\documentclass[conference,compsoc]{IEEEtran}
\IEEEoverridecommandlockouts
\usepackage{cite}
\usepackage{amsmath,amssymb,amsfonts}
\usepackage{algorithmic}
\usepackage{graphicx}
\graphicspath{{data/}{images/}}
\usepackage{textcomp}
\usepackage{listings}  
\usepackage{todonotes}
\usepackage{tabularx}
\usepackage{comment}
\usepackage{dirtytalk}
\usepackage{xcolor}
\usepackage{caption}
\usepackage{subcaption}
\usepackage{soul}
\usepackage{listings}

\colorlet{punct}{red!60!black}
\definecolor{background}{HTML}{F9F9F9}
\definecolor{delim}{RGB}{20,105,176}
\colorlet{numb}{magenta!60!black}
\definecolor{codegreen}{rgb}{0,0.6,0}
\definecolor{codegray}{rgb}{0.5,0.5,0.5}
\definecolor{codepurple}{rgb}{0.58,0,0.82}

\lstdefinelanguage{json}{
    basicstyle=\ttfamily\footnotesize,
    commentstyle=\color{codegreen},
    stringstyle=\color{codepurple},
    numberstyle=\tiny\color{codegray},
    stepnumber=1,
    numbersep=5pt,
    showstringspaces=false,
    breaklines=true,
    captionpos=b,
    backgroundcolor=\color{background},
    showtabs=false,
    showspaces=false,
    literate=
     *{0}{{{\color{numb}0}}}{1}
      {1}{{{\color{numb}1}}}{1}
      {2}{{{\color{numb}2}}}{1}
      {3}{{{\color{numb}3}}}{1}
      {4}{{{\color{numb}4}}}{1}
      {5}{{{\color{numb}5}}}{1}
      {6}{{{\color{numb}6}}}{1}
      {7}{{{\color{numb}7}}}{1}
      {8}{{{\color{numb}8}}}{1}
      {9}{{{\color{numb}9}}}{1}
      {:}{{{\color{punct}{:}}}}{1}
      {,}{{{\color{punct}{,}}}}{1}
      {\{}{{{\color{delim}{\{}}}}{1}
      {\}}{{{\color{delim}{\}}}}}{1}
      {[}{{{\color{delim}{[}}}}{1}
      {]}{{{\color{delim}{]}}}}{1}
      {//}{{{\color{codegreen}{//}}}}{1}
}

\makeatletter
\def\ScaleIfNeeded{%
\ifdim\Gin@nat@width>\linewidth
\linewidth
\else
\Gin@nat@width
\fi
}
\makeatother

\begin{document}


\title{Multi-factor authentication for users of non-internet based applications of blockchain-based platforms}

\author{\IEEEauthorblockN{Andrew Kinai, Fred Otieno, Nelson Bore, Komminist Weldemariam} 
\IEEEauthorblockA{
andkinai@ke.ibm.com, fred.otieno@ibm.com, nelsonbo@ke.ibm.com, k.weldemariam@ke.ibm.com \\
\textit{IBM Research $|$ Africa} \\
  Nairobi, Kenya 
  }


}


\maketitle

\begin{abstract}
Attacks targeting several millions of non-internet based application users are on the rise. These applications such as SMS and USSD typically do not benefit from existing multi-factor authentication methods due to the nature of their interaction interfaces and mode of operations. To address this problem, we propose an approach that augments blockchain with multi-factor authentication based on evidence from blockchain transactions combined with risk analysis. A profile of how a user performs transactions is built overtime and is used to analyse the risk level of each new transaction. If a transaction is flagged as high risk, we generate n-factor layers of authentication using past endorsed blockchain transactions. A demonstration of how we used the proposed approach to authenticate critical financial transactions in a blockchain-based asset financing  platform is also discussed.
\end{abstract}


\section{Introduction}
\label{sec:introduction}
There has been a rise in the use of non-internet based applications (e.g., SMS and USSD) particularly in the financial service industry. We call the users of these applications \emph{offline} users. First and foremost, this is driven by the large adoption of mobile money often deployed as an SMS/USSD application. As of 2018, these \emph{offline} users for mobile money reached about 866 million subscribers \cite{mobilemoney2019}. The other driving factor is the fact that only 39\% of users in low resourced regions (e.g., Sub-Saharan Africa) have smart phones \cite{mobileeconomy2019,8286115}. This means that to be able to reach more users, SMS/USSD based applications are the way to go as they can run on any mobile device. 

SMS and USSD applications primarily operate using text-based interfaces. SMS applications are linked to a phone number which a user sends and receives text messages from. For example, a user sends a text message to the SMS application linked phone number, the application performs a certain action and responds with a text message with the results of the action. Other applications operate in a conversational manner where the user and the application exchange multiple text messages before an action is executed. USSD applications on the other hand, involve dialling of a short code number on the phone. This brings up a menu where the user can select the action they would like to perform. It can also provide a text box where the user can enter values (e.g., amounts for financial applications) that can used by as an input for performing actions. 

Like users of internet-based systems (the \emph{online} users), malicious acts to gain unauthorized access to critical data (e.g., financial transactions) of \emph{offline} users has steadily increased \cite{nyamtiga2013security, yoo2015case, ACMCommunication}. However, for the \emph{offline} users,  multi-factor authentication methods are limited due to the nature of the interaction interfaces (text-based) and mode of operation of these SMS/USSD based applications. Also, from our experiences in developing and deploying several of these applications over the past years, two consistent lessons we have learned are: (i) due to the nature of interaction (text-based), these \emph{offline} users have to undertake several steps before being able to transact using such systems causing fatigue to them; and (ii) the interaction sequences involved in these applications always follow predefined sequences of steps which make it difficult to distinguish valid transaction requests from invalid ones. By exploiting these sequences,  attackers may follow a multi-stage threat workflow to break into an SMS or USSD based application with the goal of harvesting critical data.

SMS/USSD applications provide an interface for \emph{offline} users to interact and perform transactions on a system. Such an interface can be integrated with blockchain-based platforms to provide users with the flexibility to access services (e.g. financial transactions) using basic mobile phones. Platforms of this nature largely depend on the usefulness of blockchain  in handling sensitive transactions while improving trust, transparency and integrity among the participants \cite{DBLP:journals/corr/abs-1801-10228},\cite{HYPERLEDGER}. These benefits has seen the rise of integration of  SMS/USSD applications with blockchain-based solutions which still become subject to the issues and limitations that affect SMS/USSD applications.  Based on our experience of deploying a blockchain-based system integrated with an SMS application, we present our work on how we resolved these issues and limitations. 

In this paper, we present an approach aimed at augmenting blockchain with multi-factor authentication to improve the authentication experience for users of non-internet based applications while minimizing the tedious user interactions.  We evaluate our prototype implementation by using a deployed blockchain-based asset financing platform that uses an SMS application to perform critical financial transactions. 

\section{Related Work}
\label{sec:background}
Cryptography based techniques that require to update GSM infrastructure that supports SMS communication for improved SMS security has been proposed \cite{gligoric2012application, albuja2009trusted, agoyi2010sms}. Work such as \cite{nyamtiga2013security} specifically evaluated SMS/USSD mobile banking applications and recommended updating the GSM technology that supports these modes of interactions.
The same authors discussed some security weaknesses of SMS/USSD where the data sent from these applications intercepted and spoofed by attackers. Our system  addresses these issues by leveraging details of previously endorsed and committed transactions on the blockchain to verify the validity of each of these user actions. 

Use of various combinations of authentication technologies has been proposed to increase the security of systems. The user-friendliness of various combinations of authentication technologies been evaluated in  \cite{banyal2013multi, kim2011method}
The use of graphical passwords for authentication has also been widely explored \cite{sabzevar2008universal,BiddleCO12}. However, all of these mainly focused on internet based applications and some of these approaches cannot be applied to SMS/USSD based applications. 

A desired requirement to secure systems is being able to detect unauthorized access in real-time. In line with this, several approaches on detecting automated bots masquerading as valid users by analyzing their actions in the systems \cite{lee2013know} and 
anomaly behaviour detection \cite{villamarin2008identifying}
have been the subject of computer security over several decades.  CAPTCHA \cite{CAPTCHA}, an automated test that (non)human users can pass, is one of such approaches applied to verify/deny suspicious requests \cite{von2003captcha}. 
These technologies focus on mainly internet-based applications. There are many applications that use SMS/USSD as a point of interaction for users. SMS/USSD is an insecure channel that can be intercepted and spoofed by attackers \cite{nyamtiga2013security}. Our approach is designed to address this issue of SMS/USSD based application by leveraging details of the current transaction to verify whether the origin of the SMS/USSD command is actually the valid user meant to undertake the transaction.

Another important feature of a secure system is its ability to detect fraudulent transactions. Fraud detection in credit card and e-commerce transactions has been explored in \cite{bolton2002statistical, chan1999distributed, dorronsoro1997neural} using machine learning techniques. They used labeled past transactions (fraudulent and non-fraudulent) to train machine learning models that are able to flag new fraudulent transactions when they are carried out. Our approach is an enhancement to these methods by training a specific machine model for each user. This ensures fraudulent detection of transactions is tailored to how a specific user performs valid transactions not how all users in general perform valid transactions.

\section{Approach}
\label{sec:proposedframework}

 We introduce a seamless layer of authentication management for blockchain-based applications that use SMS/USSD as a mode of interaction. This authentication layer augments a  multi-factor authentication framework using past blockchain transactions to generate challenge questions and answers.
%
%
We define authentication requirements that use smart contracts and blockchain system for two objectives. First, we focus on detecting any suspicious transaction request originated from SMS/USSD based applications.  Secondly, the ability to generate contextual authentication mechanisms in order to validate suspicious transaction requests. 
\begin{figure}[h!]
	\centering
		\includegraphics[width=1\ScaleIfNeeded]{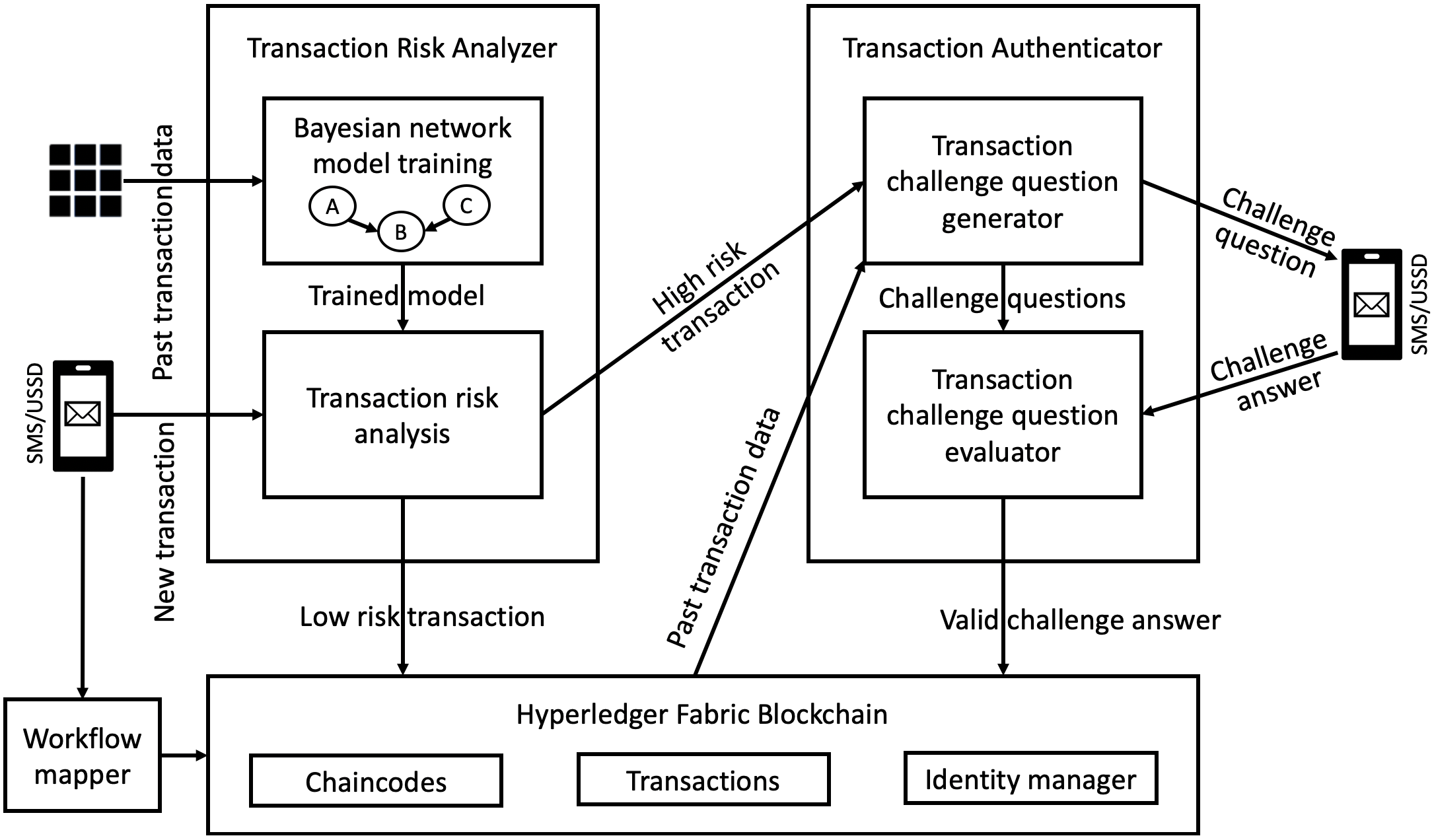}
	\caption{Overview of our approach}
	\label{fig:approach}
\end{figure}
Figure \ref{fig:approach} shows the schematic overview of our approach. Below 
we focus on three key aspects of our proposed system: workflow generation from non-internet based applications by parsing SMS message sequences of user interaction, transaction risk analysis based on a user's profile, and multi-factor authentication using previously endorsed transactions.  

\subsection{From "Offline" transactions to Smart Contracts}
\label{subsec.workflow}

By synthesizing and modeling the interaction/message sequences (\emph{offline transactions}) between a user and a given SMS/USSD application, we map these sequences into corresponding workflows (a collection of smart contracts) on a blockchain system.  The resulting workflows define the participants of a transaction, sequence of actions carried out to perform the transaction, and the participants that carry out each action.  This is done by the Workflow Mapper as shown in Figure \ref{fig:approach}.

Our Workflow Mapper is inspired by, and builds upon, the 
approach presented in \cite{bhattacharya2007towards, hull2010introducing}. The approach focuses on identifying the entities that are involved in a business process transaction that have a distinct life cycle. In particular, we built on their artifact-centric business  process methodology and formalism for business process mapping of a given SMS/USSD application into a collection of workflows.
The mapping of an SMS/USSD application into a transaction  workflow is a one time operation occurs in two steps:  
\paragraph{Transaction life cycle steps  identification} For a given SMS/USSD application, we analyze a series of text messages a user needs to send and receive to perform a valid transaction using a custom trained Natural Language Processing (NLP) model. The analysis generates possible sequence of steps/actions (that model a valid transaction lifecycle), which are then mapped into workflows on blockchain.
We then use an instance of the generated workflow to track all the user interaction events, which are persisted on the blockchain. 
\paragraph{Transaction life cycle step challenge template generation} 

For each identified transaction workflow step in Step a) above, we generate CAPTCHA like candidate challenge templates (as $<$question, answer$>$ pairs). To do so, first, using part-of-speech tagging NLP techniques, we identify the different types of transaction steps (e.g., option selection, confirmation, etc.).  Secondly, we define challenge template for each identified type of the workflow transaction step or reuse from previously defined templates by performing transaction similarity analysis.  We also store the generated templates in the blockchain together with the associated transaction workflow. Finally, the templates will be instantiated when an instance of a transaction is being executed in real-time. 
An example of a challenge template for an option selection transaction step of the first SMS in Figure \ref{fig:sms_application} is:   

$challengeTemplate_i:$ $Pay$ $[answer_1]$ $in$ $[answer_2]$ $days;$  where $answer_1$ and $answer_2$ represent candidate  challenge answers of the template that can be set as blanks when generating the fill-in-the-blank challenge question.
\subsection{Transaction risk analysis} 



Every time a user performs an SMS/USSD transaction, the Transaction Risk Analyzer determines the legitimacy/anomalousness of the transaction using a Bayesian Network (BN). BN is a probabilistic graphical model with characteristics enabling representation of conditional dependencies in transactions\textquotesingle \space features. This is useful in predicting the likelihood of a transaction being fraudulent. We used historical transaction logs collected from real-world application over the course of 2 months pilot to train and bootstrap the initial BN model, in which nodes represent features (specific characteristics) of transactions, and the directed edges (arcs) represent the conditional dependencies between these features of transactions.

In particular, from the logs and for each user, we extracted relevant features such as total transaction time, time taken between transaction steps, errors committed when performing transaction actions. The BN model rates 
each incoming transaction against the \emph{offline} user transaction profile model to identify any anomalous behavior of the transaction. 
Examples of anomalous behaviors include unusually high number of action requests in a short period of time,  unusually high number of multiple invalid action requests which could be originating from automated bots, transactions performed in a time period the user does not normally perform transactions etc. 
Note that we model, as a conditional dependence on the BN model, the \emph{offline} user profile using a combination of their interaction events,  basic user metadata (e.g., demographic data, gender, business profile, etc.), and transaction features/characteristics (e.g., total transaction time, time taken between transactions, average value of transactions, etc.).

The conditional dependence is important because some features of a transaction that are used to develop a user's transaction profile may have a causal effect/influence toward other features of a transaction. For example, in an goods ordering application for retailers, the time between transactions may influence the value of the next transaction (i.e. the longer the retailer stays without making an order, the higher the probability that their next order will be of a high value).
These variables and their conditional dependence may differ for different SMS/USSD applications. 
The structure of the BN is trained by analysing the data generated by the SMS/USSD application to identify the feature variables and their relationships. The probability tables for each user are bootstrapped using past user transaction data. When a new transaction action is received by the system, the transaction profile is evaluated against the profile of that user to determine the risk level of the transaction.
If the transaction under analysis identified to be suspicious, the Transaction Authenticator will be triggered, and the relevant events are logged in to the ledger. Otherwise, if the transaction action matches the user's transaction profile model, the transaction action is accepted as valid and stored in the ledger.
\subsection{Multi-factor authentication} 
Once the incoming transaction is analyzed and its risk level is determined to be high,  the Transaction Authenticator triggers multi-factor authentication by executing one or more predefined rules. 
The risk level of the incoming transaction under analysis determines the rules that are to be executed.
Below is an example of a rule:

$rule_i:$ $if(isRisk(tx) \geqslant  multiFactorThreshold) \rightarrow mAuth= newMultiFactorAuth(numberOfQuestions);$

%
%
%

\noindent where 
\emph{multiFactorThreshold} defines the risk level threshold for the transaction ($tx$) based on which the Transaction Authenticator determines a multi-factor authentication at run-time (in this case a CAPTCHA like challenges) based on past confirmed transactions. 
This value is determined at run-time based on the number of past successful transactions a user has performed. The fewer the number of past transactions a user has, the weaker their transaction profile model that is used to identify the risk level of a transaction. Therefore, a higher value of  \emph{multiFactorThreshold} will be set to reduce false positives when the user's transactions are being analysed. This value will be reduced as the user continues to perform more transactions in the system and their transaction profile model becomes better overtime.

In the example $rule_i$ above, the \emph{numberOfQuestions} defines the number of challenge questions that should be generated once multi-factor authentication is triggered. In fact, this depends on the number of past transactions available (on the ledger) to be used to generate the questions and the number of chances the user is to get to validate their transaction. 
%
%
This set of challenges is dynamically added  to  the  workflow  of  this  specific transaction following Step b) in Section \ref{subsec.workflow}.  This  adds new actions in the transaction workflow which the user has to respond to correctly for the transaction to proceed to the next normal execution step.
If the user is unable to answer the first multi-factor authentication challenge correctly the Transaction Authenticator will be triggered to generate the next set of challenges using updated risk level (from the BN model) as an additional input. 
\section{Implementation and Illustration}

\subsection{Use case scenario}
\begin{figure}[h!]
	\centering
		\includegraphics[width=0.75\ScaleIfNeeded]{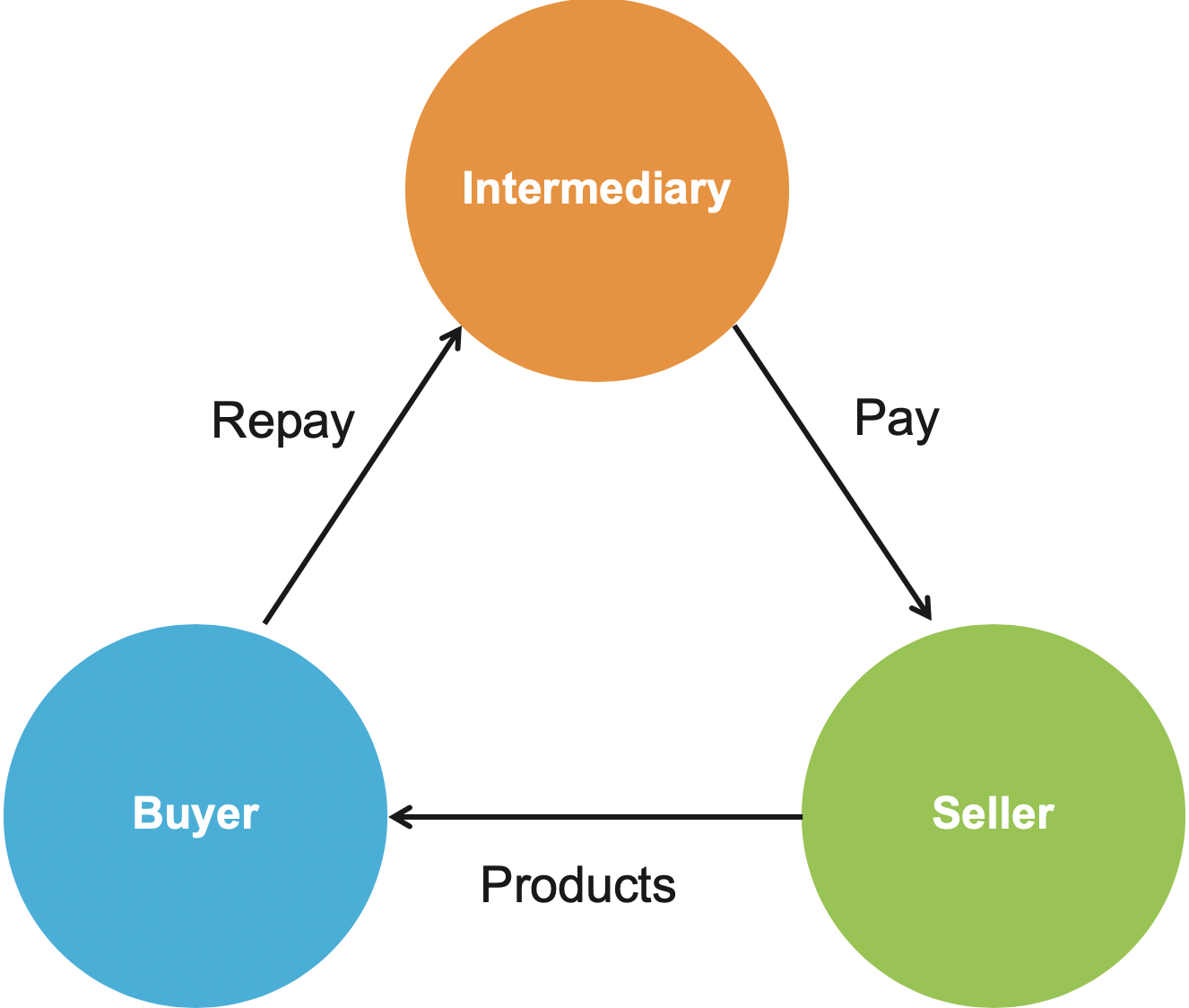}
	\caption{3-party transaction model of the blockchain-based asset financing platform}
	\label{fig:participants}
\end{figure}

To illustrate our approach, consider a blockchain-based asset financing platform that facilitates ordering  of  goods  from  distributors and financing of the orders by financial institutions \cite{kinai2017asset}. The platform is based on a 3-party transaction model to facilitate financing at the point of purchase as shown in Figure \ref{fig:participants}. The 3 stakeholders in the model are: \emph{(i) buyer}: a customer requesting to purchase goods or services, \emph{(ii) seller}: an entity that offers a set of goods or services at a given price and \emph{(iii) intermediary}: an entity that is legally entitled to provide financing for a set of goods or services. 

Each buyer in the platform has a credit profile that is computed by running a machine learning model on their past transactions. The buyer's credit profile is composed of a credit score and a credit limit. The \emph{credit score} is used to determine the interest rate/service fee charged to a buyer when they take a loan to finance a new order. The \emph{credit limit} determines the maximum loan amount that can be advanced to a buyer.  

The buyer's credit profile is what the intermediary uses to determine where of not to finance an order a buyer makes to a seller and it is important to ensure the correct data is used to compute it. To facilitate secure and traceable sharing of data between these stakeholders, the platform is implemented on the Hyperledger Fabric \cite{HYPERLEDGER} blockchain. The initial setup of the blockchain network is  configured and deployed for 3 organization clusters (buyers, intermediary, sellers) each with 1 peer. The organization clusters (orgs) are subscribed to a common channel. The platform is scalable and can support multiple sellers ensuring privacy of data between sellers by using a separate channel for each seller.

The pilot deployment of the platform was done as an SMS application because majority of the traders participating in the pilot did not have access to smart phones. 

The process of executing a transaction in the platform using the SMS application is as follows. 
\begin{enumerate}
  \item A distributor's sales representative (seller) confirms an order of goods a trader would like to buy and the trader (buyer) receives a confirmation SMS.
  \item If the trader meets the criteria to be offered financing for the order by a financial institution (intermediary), they get loan  offers via SMS.
  \item The trader responds via SMS with the loan term offer they like.
  \item The financing of the order is confirmed and the trader receives an SMS with details on how to pay back the loan.
\end{enumerate}

\begin{figure}[h!]
	\centering
		\includegraphics[width=0.5\ScaleIfNeeded]{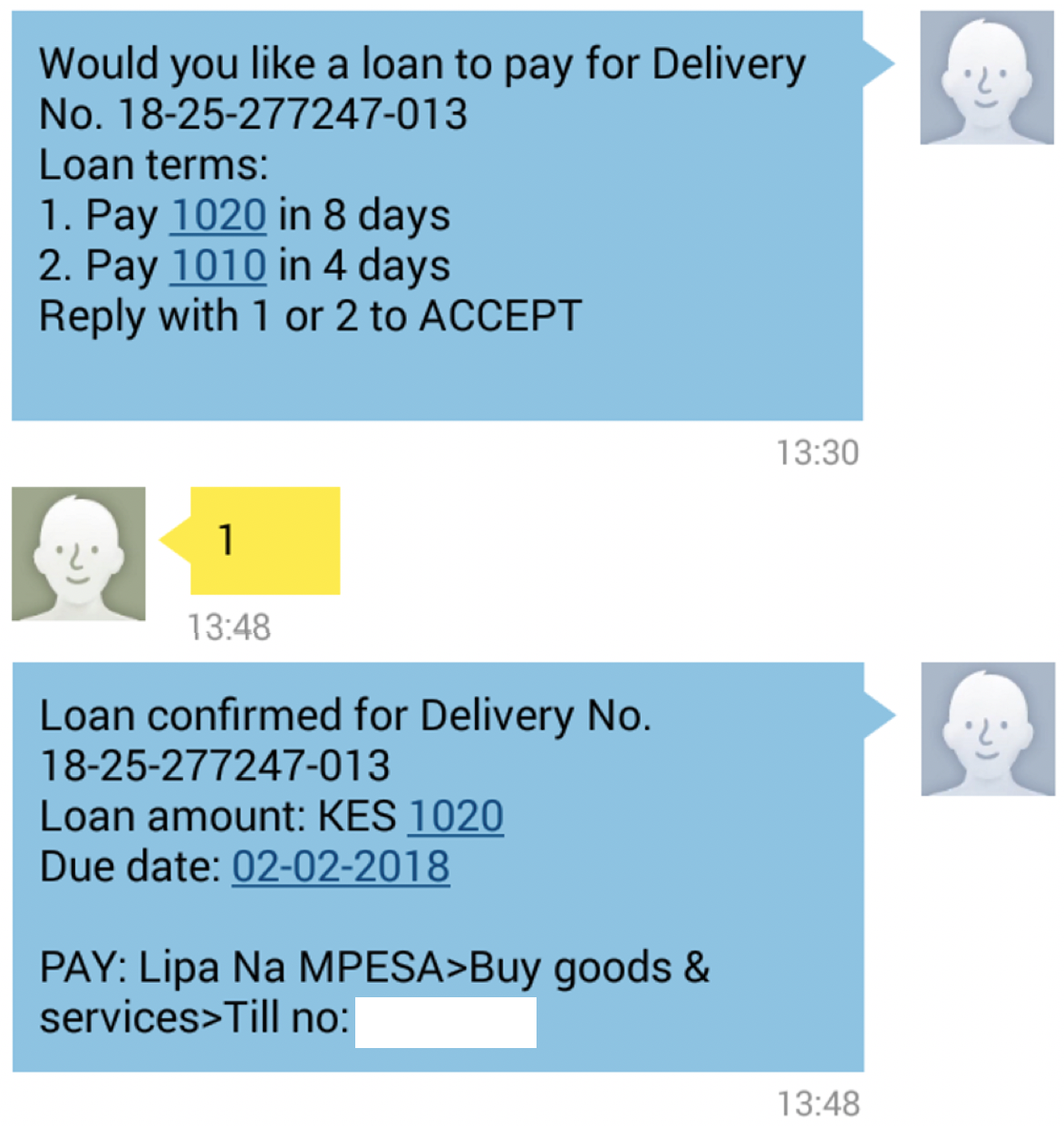}
	\caption{Example SMS message sequences for a buyer to confirm loan financing for an order}
	\label{fig:sms_application}
\end{figure}

Figure \ref{fig:sms_application} shows an example of a sequence of SMS messages required to execute an order transaction in the blockchain-based asset financing platform (excluding the order confirmation SMS for Step 1) via the SMS application. The first SMS message shows the loan options offered to the buyer for their order by the financier. The second SMS message shows the buyer sending "1" to indicate they have selected the first loan option. The third SMS message is from the platform confirming the loan and indicating the loan amount and the loan's due date.

The smart phone applications of the platform provided the authentication mechanisms required for secure transactions in the platform that were not available for the SMS application. The smart phone applications supported authentication of users using usernames and passwords and use of a user token to authorize all blockchain transactions. In the case of the SMS application, token management possible on the client side. This would lead the user having to authenticate themselves every time they send an SMS command to the system. This repeated authentication would lead to user fatigue and thus motivating this work.

\subsection{Implementation details}
\noindent 

The multi-factor authentication system is implemented as a suite of cloud foundry micro-services and chaincodes are implemented on Hyperledger Fabric \cite{HYPERLEDGER}. It is deployed in the same blockchain network of the lending platform leveraging the same org, peer and channel configurations. For the machine learning model, we used the Trusted Model Executor \cite{abs-1911-03623},  which is designed to  execute, evaluate, and track the usage and performance of machine learning models on top of blockchain.
%



During run-time, we authenticate SMS messages in the following way. First, all the SMS messages from users are received and intercepted. Second, every SMS received contains user mobile number and the content of the message, upon receiving this message we retrieve the current state of the workflow from the blockchain.  
We further retrieve all transaction actions (current and previous)  of the ongoing transaction workflow from the ledger. Examples of (for the ordering use case) these transaction actions include order confirmation (challenge answer:  value of goods ordered), loan terms offered (challenge answer: loan amount), etc.  The hash of a challenge answer together with its corresponding transaction action is stored in the ledger. For example, when an order is confirmed, the hash value of the order amount is stored which can be used to generate a question \say{What is the value of your order?}. The hash will be used for counter-checking integrity and correctness of the user\textquotesingle s response to a posed challenge. Third, three or more of the last completed transactions from the ledger will be selected to generate challenge questions and answers. These challenge questions will be embedded in the transaction workflow and updated in the ledger. 


An example of the mapped \textit{OrderConfirmation} workflow event for a transaction where an order placed by a \emph{trader} and confirmed by a \emph{distributor} 
is shown in Listing \ref{workflow1}. This event contains the hash of the value of the order.

\begin{lstlisting}[label=workflow1, language=json, basicstyle=\tiny, caption=Example of a high-level view of workflow events after order confirmation]
{
    "events": [
        {
            "doc0": {
                "user": "SalesRepresentative",
                "organization": "Distributor",
                "eventType": "OrderConfirmation",
                "docHash": "5E8FF9BF55B" // Contains the hashed order amount of the confirmed order
            }
        }
    ]
}
\end{lstlisting}

\begin{figure}[h!]
	\centering
		\includegraphics[width=1\ScaleIfNeeded]{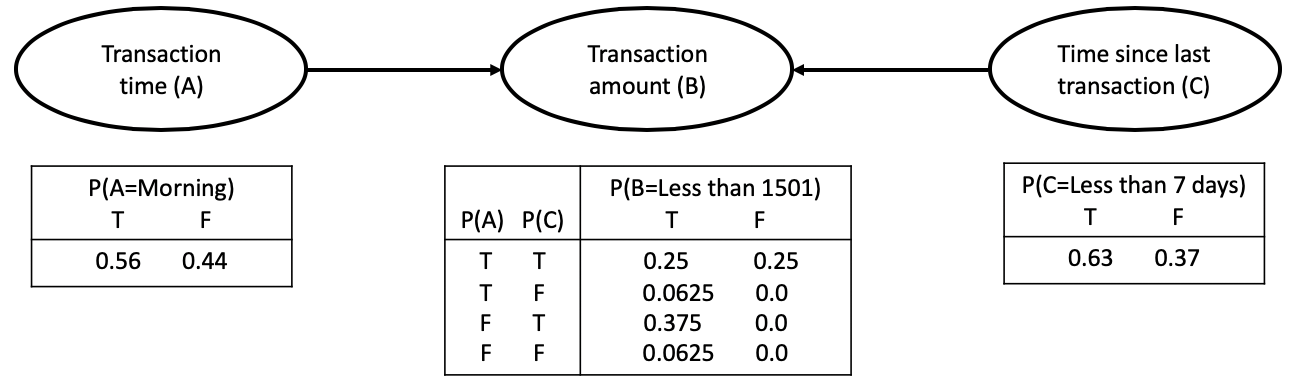}
	\caption{A simple example illustrating a BN to assess the risk level of a transaction and probability tables of one user.}
	\label{fig:bayesian_network}
\end{figure}

We trained a simple BN model shown in Figure \ref{fig:bayesian_network}. As illustration, we only used three feature variables for the network: transaction time (A), transaction amount (B), and time since the last transaction (C). The probability tables for each user are bootstrapped using past user transaction data. We evaluated 200 valid transactions from a real-world pilot deployment of our SMS-based lending platform \cite{kinai2017asset} to derive these feature variables (and after consulting with the domain experts from a bank). Transaction time (A) is the time of day that the transaction occurred. There were 2 classes identified from the data: morning (6.00 am to 11.59 am) and afternoon (12.00 pm to 18.00 pm). Transaction amount (B) is value of a transaction. The maximum value of transactions during the pilot was KES 3000 and this was divided into 2 classes: amounts less than KES 1501 and amounts greater than or equal to KES 1501. Time since last transaction (C) represents the time that has elapsed since the last transaction performed by a user. This was divided into two classes: less than 7 days or greater than or equal to 7 days. From the transactions, we observed that the transaction amount depends on the transaction time and the time since the last transaction as shown in  Figure \ref{fig:bayesian_network}. The figure also shows the probability tables that represent a specific (\emph{offline}) user in the pilot. We can see this user does transaction valued at less than KES 1501 75\% of the times.


\begin{figure*}[h!]
	\centering
		\includegraphics[width=1\ScaleIfNeeded]{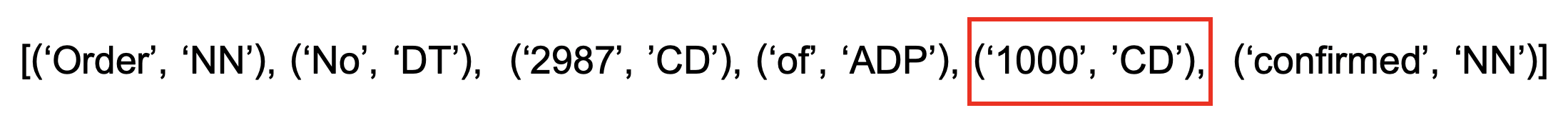}
	\caption{An example illustrating how part-of-speech tagging is used to generate challenge questions and answers}
	\label{fig:sentence_tagging}
\end{figure*}

\begin{lstlisting}[label=workflow2,language=json, basicstyle=\tiny, caption=Example of a high-level view of workflow events after augmented authentication]
{
    "events": [
        {
            "doc0": {
                "user": "SalesRepresentative",
                "organization": "Distributor",
                "eventType": "OrderConfirmation",
                "docHash": "5E8FF9BF55B" // Contains the hashed order amount of the confirmed order
            }
        },
        {
            "doc1": {
                "user": "Shopkeeper",
                "organization": "SME",
                "eventType": "AugmentedAuthentication",
                "docHash": "B015CFEC81A" // Contains the hashed challenge correctly answered derived from the previous action
            }
        }
    ]
}
\end{lstlisting}

After assessment of the order transaction using the BN shown in Figure \ref{fig:bayesian_network}, the transaction risk level is determined to be higher than the \emph{multiFactorThreshold}. A challenge fill-in-the-blank question is dynamically generated based on the order value amount (using the template as in Section \ref{subsec.workflow}) and sent to the user via SMS. We perform part-of-speech tagging on the previous order confirmation text to generate the challenge question and answer. This entails breaking down the sentence into its part-of-speech components and generating the challenge question and answer as shown in Figure \ref{fig:sentence_tagging}. From the confirmation text \say{Order No 2987 of 1000 confirmed}, we generated the challenge: \say{What was the amount of your last order: Order No 2987 of  \_\_\_ confirmed}. 
If the user sends a correct answer back to the system via SMS, the \textit{AugmentedAuthentication} event is added to the workflow as shown in the second event in Listing \ref{workflow2}.

The user is now be able to proceed to the next workflow action which is requesting for a loan for the order. The new event \textit{RequestForLoan} will then be added to the workflow as shown in Listing \ref{workflow3}.
The user will then be able to continue transacting in the system following the remaining workflow steps as defined in the original transaction workflow.

\begin{lstlisting}[label=workflow3,language=json, basicstyle=\tiny, caption=Example of high-level view of workflow events]
{
    "events": [
        {
            "doc0": {
                "user": "SalesRepresentative",
                "organization": "Distributor",
                "eventType": "OrderConfirmation",
                "docHash": "5E8FF9BF55B" // Contains the hashed order amount of the confirmed order
            }
        },
        {
            "doc1": {
                "user": "Shopkeeper",
                "organization": "SME",
                "eventType": "AugmentedAuthentication",
                "docHash": "B015CFEC81A" // Contains the hashed challenge correctly answered derived from the previous action
            }
        },
        {
            "doc2": {
                "user": "Shopkeeper",
                "organization": "SME",
                "eventType": "RequestForLoan",
                "docHash": "44RER68j5E8" // Contains the hashed loan amount of the confirmed order
            }
        }
    ]
}
\end{lstlisting}

\section{Discussion and Conclusion}
\label{sec:conclusion}
In this paper, we have presented the design and implementation of our initial approach in addressing a recurrent challenge we have been experiencing (how to effectively authenticate \emph{offline} users who rely on non-internet based applications to prevent unauthorized access to their valuable information).  We proposed to augment blockchain with multi-factor authentication for these users 
by mapping the operations of such applications (SMS/USSD) into blockchain-enabled workflows. Hyperledger fabric is likely to be affected by large transactions, detailed performance optimization and guidelines are given in  \cite{DBLP:journals/corr/abs-1801-10228,DBLP:journals/corr/abs-1901-00910} for various deployment configurations which are out of scope for this paper. For each mapped transaction workflow step, we automatically generated a set of CAPTCHA like challenges. 

We trained and evaluated a simplistic Bayesian network model (developed using historical transaction logs) for detecting malicious transactions at run-time.  We then presented an approach to determine multi-factor authentication for a user that would trigger the execution of multi-factor authentication workflow accordingly.  One of the main benefits of our approach is that it does not require to instrument the SMS and USSD application.
The key insight in our approach is to leverage the execution steps of the SMS/USSD application and map them to workflows. This will reduce the number of times a user authenticate by entering a password. The transaction risk analysis will be used to determine whether the transaction is valid instead of the user always authenticates themselves every time they use the SMS/USSD application.

While our initial approach gained attraction from our banking partners; however, we need to evaluate the effectiveness of our approach using different SMS/USSD based applications and historical transaction logs. Hence, we plan to conduct extensive pilot experiments. As part of the pilot experiments, we will evaluate the transaction risk level analysis with the transaction data from the SMS-based lending application deployment by comparing different users with different number of past transactions to be able to come up with a function that defines the relationship between the number of transactions and the \emph{multiFactorThreshold} assigned to a user during run-time. 
\section{Future Work}
We are exploring novel approaches to enable users using either SMS or USSD to directly transact and participate in a blockchain network. Some of these approaches include: i) using a mobile network as a blockchain node that users in the network can be authenticated while reducing cross-network latency; ii) using SIM-based applications that are embedded on the users SIM-card with any extra information needed for trusted communication with a mobile provider network. This will ultimately allow non-internet based application users to join and enjoy the value of blockchain-based solutions.

\bibliography{bibliography}
\bibliographystyle{IEEEtran}

\end{document}